\documentclass[pdflatex,sn-mathphys-ay]{sn-jnl}

\usepackage{xurl}
\usepackage{hyperref}
\usepackage{graphicx}%
\usepackage{multirow}%
\usepackage{amsmath,amssymb,amsfonts}%
\usepackage{amsthm}%
\usepackage{mathrsfs}%
\usepackage[title]{appendix}%
\usepackage{xcolor}%
\usepackage{textcomp}%
\usepackage{manyfoot}%
\usepackage{booktabs}%
\usepackage{algorithm}%
\usepackage{algorithmicx}%
\usepackage{algpseudocode}%
\usepackage{listings}%
\usepackage{orcidlink} 
\usepackage{natbib}

\begin{document}

\title[Article Title]{Reference Coverage Analysis of OpenAlex compared to Web of Science and Scopus}

\author[1, *]{\fnm{Jack H.} \sur{Culbert}\orcidlink{0009-0000-1581-4021}}
\author[2]{\fnm{Anne} \sur{Hobert}\orcidlink{0000-0003-2429-2995}}
\author[2]{\fnm{Najko} \sur{Jahn}\orcidlink{0000-0001-5105-1463}}
\author[2]{\fnm{Nick} \sur{Haupka}\orcidlink{0009-0002-6478-6789}}
\author[3]{\fnm{Marion} \sur{Schmidt}\orcidlink{0000-0001-6970-3714}}
\author[3]{\fnm{Paul} \sur{Donner}\orcidlink{0000-0001-5737-8483}}
\author[1]{\fnm{Philipp} \sur{Mayr}\orcidlink{0000-0002-6656-1658}
}

\affil[1]{\orgname{GESIS -- Leibniz Institute for the Social Sciences}, 
\country{Germany}}

\affil[2]{\orgdiv{Nieders\"achsische Staats- und Universit\"atsbibliothek G\"ottingen}, \orgname{Georg-August-Universit\"at}, 
\city{G\"ottingen}, 
\country{Germany}}

\affil[3] {\orgname{Deutsches Zentrum f\"ur Hochschul- und Wissenschaftsforschung}, 
\city{Berlin}, 
\country{Germany}}
\affil[*]{Corresponding Author: Jack H. Culbert, \href{mailto:jack.culbert@gesis.org}{jack.culbert@gesis.org}}
\abstract{OpenAlex is a promising open source of scholarly metadata, and competitor to established proprietary sources, such as the Web of Science and Scopus.
As \mbox{OpenAlex} provides its data freely and openly, it permits researchers to perform bibliometric studies that can be reproduced in the community without licensing barriers. However, as OpenAlex is a rapidly evolving source and the data contained within is expanding and also quickly changing, the question naturally arises as to the trustworthiness of its data. 
In this report, we will study the reference coverage and selected metadata within each database and compare them with each other to help address this open question in bibliometrics. In our large-scale study, we demonstrate that, when restricted to a cleaned dataset of 16.8 million recent publications shared by all three databases, \mbox{OpenAlex} has average source reference numbers and internal coverage rates comparable to both Web of Science and Scopus. We further analyse the metadata in OpenAlex, the Web of Science and Scopus by journal, finding a similarity in the distribution of source reference counts in the Web of Science and Scopus as compared to OpenAlex. We also demonstrate that the comparison of other core metadata covered by \mbox{OpenAlex} shows mixed results when broken down by journal, capturing more ORCID identifiers, fewer abstracts and a similar number of Open Access status indicators per article when compared to both the Web of Science and Scopus.}




\maketitle

\section{Introduction}\label{sec1}

OpenAlex \citep{priem2022} was released on January 1st 2022 by OurResearch as a replacement for the discontinued Microsoft Academic Graph (MAG) and is offered as a fully open source of scholarly metadata, with all data, API information and code released to the public. 
As observed in the comparative study by \cite{scheidsteger2022}, not all aspects of the MAG were reproduced, as patents were not captured in OpenAlex. Aside from this exception, OpenAlex is effectively a continuation and expansion of the MAG.

OpenAlex is a promising alternative to proprietary bibliometric data sources as its permissible licensing creates the potential to support a transformation of research practice towards reproducible bibliometrics. This is being realised in open research policies in academia, for example in December 2023,  Sorbonne University has switched from using the Web of Science (WoS) and Clarivate bibliometric tools to OpenAlex and open-source tools.\footnote{\url{https://www.sorbonne-universite.fr/en/news/sorbonne-university-unsubscribes-web-science}} Reproducible bibliometric research is hardly possible with proprietary bibliometric data sources as their licensing terms rule out dissemination of data.

As a widely used open source repository of scholarly metadata, OpenAlex has previously been the subject of research as to its suitability for a variety of bibliometric analyses, such as the review by \cite{velez-estevez2023}, which comparatively analysed various APIs to bibliometric corpora, including API interoperability, characteristics and their use in research practice,  and \cite{RePEc:dem:wpaper:wp-2023-018} who released a working paper on the migration of scholars which included a comparative study between Scopus and OpenAlex, limited to the coverage of scholars in Western and non-Western countries.

Although country affiliation metadata accuracy and completeness were found lacking in earlier versions \cite{zhang_missing_2024}, OpenAlex was recently considered suitable for countrywide analyses by \cite{alperin_analysis_2024}. Investigating diamond open access journals indexed in OpenAlex in comparision with both WoS and Scopus, \cite{simard_open_2024} highlighted that OpenAlex journal indexing is more inclusive than that of WoS and Scopus.
However, at this early stage of its development, OpenAlex is a highly dynamic data source whose characteristics change with each release. 
This paper also follows previous quantitative comparisons of citation coverage of traditional bibliometric databases. One such study  examines a dataset of 2.5 thousand documents published in 2006 and approximately 3.1 million citations of these documents, found in the MAG, Google Scholar, WoS, Scopus and OpenCitations' COCI databases, \citep{martin-martin_google_2021}. This study demonstrates relative coverage gaps in some subject areas in the MAG as compared to WoS and Scopus, and also overall that Google Scholar has the largest citation coverage as compared to the other databases.

 This concurs with an earlier study also by \cite{martin-martin_google_2018},  which also compared these databases to Google Scholar, where approximately 2.45 million citations from 2300 documents covering 252 subjects are compared for similarities in the citation coverage, similarly finding the citations in Google Scholar comprises a superset of those in WoS and Scopus.

Previous work has studied and compared bibliographic databases to better understand the limitations of different data sources. Comparing  the MAG, Scopus, WoS and other databases, \cite{visser_large-scale_2021}, argue for combining databases to allow for comprehensive coverage, taking into account the strengths and weaknesses of the different data sources. Similarly, other studies have focused on a journal coverage analysis of WoS, Scopus and Dimensions, such as the study of \cite{singh2021}. Furthermore tools such as F\"arber's tool for comparing author records between databases \cite{farber_which_2022}, have been created and shared by the academic community to provide insights into the suitability and weaknesses of bibliometric databases for accurate bibliometrics in their current state. 

Therefore to enable bibliometricians to better understand the potentials and current limitations of OpenAlex, we compare OpenAlex with two major proprietary bibliometric data sources, WoS and Scopus. With our study, we wish to contribute to the question to what extent OpenAlex can serve as an adequate, (or even better) free alternative to established, proprietary databases for bibliometric research and reporting. Our specific research questions in this report are whether reference coverage of items differs between the three data sources, investigating this for the complete databases as well as for a sub-corpus of items present in all three databases, and whether and to what extent the coverage of some additional metadata fields, specifically abstracts, Open Researcher and Contributor IDs (ORCIDs), and Open Access status of items differ in all three data sources.

We are aware that these initial assessments are likely to change with further developments, as of writing twelve new snapshots of OpenAlex have been released - with new data added or modified regularly, so this report should be understood as reflecting the state as of late 2023. Since then at least 151 million new references have been added which was an increase of 7.61\% while at least 750,000 records were deleted and over 3.4 million records were added.\footnote{\url{https://github.com/ourresearch/openalex-guts/blob/main/files-for-datadumps/standard-format/RELEASE_NOTES.txt}}

\subsection{Reference Coverage}
References are of central importance for bibliometric databases, as matching them to their target items forms the basis for the calculation of citation metrics.
As a first step, we compare average reference counts between the three databases, whereby the basis of the comparison are the complete databases, then subsets of publications with the document type `article', and a shared sub-corpus of publications covered by all three databases. Citation reference data can also be used for an indirect assessment of the coverage, i.e. the proportion of relevant research publications that are included in the database and accessible to users for analysis \citep{singh2021}. An insufficient or biased coverage of the relevant literature should rule out the use of a database for a particular study. 

There are different ways to determine the coverage of a database, for example, the comparison with external lists of relevant sources or publication lists of a sample of representative researchers of the studied fields.  However, there is no general gold standard corpus and all external sources therefore bring their own biases and limitations. A relatively simple  (and easily replicable and repeatable) way to study literature coverage is calculating the internal reference coverage of a database as a whole or in relation to grouping characteristics, such as disciplines, the literature of particular countries or language communities. 

The internal coverage is the proportion of those cited references of a publication set which are themselves covered as source items in the database, out of all cited references in the set. We refer to these as source references and in contrast, to references to items that are not themselves indexed in the database as non-source references (or references to non-source items).

A more comprehensive introduction to this concept and an analysis of the internal coverage is available in \citet[Chapter 7]{moed2005citation} and \citet{van_raan_measuring_2019}. The great advantage of this type of analysis is that one does not need any external data which may be difficult and costly to collect. This reliance on only the assessed data source itself is also the major disadvantage, as one is limited to the reference data as present in the assessed data source with all its contingencies. Therefore one cannot simply extrapolate from the coverage of cited literature to the coverage of literature segments that were never cited in the source data, possibly as a direct consequence of the source database's selection criteria. These considerations show why internal reference coverage provides merely a partial and possibly source-biased measurement of coverage.

Nevertheless, when comparing citation index databases, the differences in internal reference collection can be a useful guide. For example, one question that arises due to the much larger dimension of OpenAlex compared to WoS and Scopus is whether it thereby also has a higher internal coverage, i.e. a higher proportion of publications that are referenced and also indexed in the database compared to the other databases.
There are no established guidelines for numerical values of coverage proportions required to allow reliable studies to be carried out. But for example, \cite{moed2005citation} analysed the combined ISI Citation Indexes (the predecessor of today's Web of Science) and found that the coverage rate, which is the proportion of references from the 2002 source year that refer to ISI source journals, was highest for Molecular Biology and Biochemistry, at around 90\%, followed by human-focused Biological Sciences, Chemistry, Clinical Medicine and Physics and Astronomy. It was vastly lower in the Arts and Humanities and intermediary in the Social Sciences, Mathematics and Engineering. 

When using this indicator to compare OpenAlex, WoS and Scopus, we are thus less interested in an evaluation in absolute values, but rather in assessing how OpenAlex performs in comparison to the two established bibliometric databases. 
Internal reference coverage depends on the size and possibly the disciplinary profile of a database as well as the accuracy of its reference matching procedure. As OpenAlex is actually much larger than Scopus and WoS (see Table \ref{tab:data-overview-size}) it could be expected that its internal reference coverage is at least not lower than those of the latter databases.  

\subsection{Open Metadata}

The increasing discussion surrounding the open availability and quality of various types of scholarly metadata in bibliometrics is not limited to reference coverage, but expands to other metadata \citep{van_Eck_2023, delgado-quiros_completeness_2024, cespedes_evaluating_2024, zhang_missing_2024}. For instance, the Initiative for Open Abstracts (I4OA)\footnote{\url{https://i4oa.org/} advocates open abstracts of scholarly works and calls on scholarly publishers to submit them to Crossref,\footnote{\url{https://www.crossref.org/}} a Digital Object Identifier (DOI) registration agency. Similarly, scholarly publishers can use Crossref to share the funding information associated with the articles they publish.} However, coverage analyses of Crossref suggest that not all publishers provide open scholarly metadata to Crossref \citep{Mugabushaka_2022,Kramer_2022}.
Another example of essential metadata is the use of ORCIDs to persistently identify authors, helping bibliometricians not only to disambiguate author names, but also to interlink different data from different sources based on the ORCID \citep{haak_2012}.

As open data sources are essential for OpenAlex, we will expand our analysis to compare abstracts, funding information and ORCID coverage at the journal level. Moreover, we will assess the coverage of open access status information between OpenAlex and the proprietary databases WoS and Scopus. In contrast to abstract and author information, all three databases use the same source, the open access discovery service Unpaywall, to retrieve open access status information \citep{Else_2018}.

\section{Data and Methodology} \label{sec-method-data}
In this section, we describe the data used in this study, and the reasoning for our choices of restrictions and subsets of this data. To enable a fair comparison between OpenAlex, and WoS and Scopus, we have created a `Shared Corpus' containing records common to all three datasets based on an exact DOI match, which have been published between 2015 and 2022, where the DOI is unique to the record in all three databases, i.e. there are no multiple records with the same DOI. In the course of  selecting records from the databases, it is ensured that publications only ever have one DOI assigned to the record. In a further step, the references of the publications in the Shared Corpus are restricted to those published between 1996 to 2022.

\begin{table}[ht]
    \centering
    \begin{tabular}{l|rrr}
         & WoS & Scopus & OpenAlex  \\
         \hline
         \hspace{-2mm}\textit{Whole Corpus} & & & \\
         Number of Records & 71,280,830 & 65,642,377 & 243,053,925 \\
         Number of References & 1,765,281,799 & 2,033,522,623 & 1,845,379,285 \\
         \\
         \hspace{-2mm}\textit{Whole Corpus - Articles Only} & & & \\
         Number of Records & 42,678,632 & 43,579,595 & 200,665,940 \\
         Number of References & 1,400,958,343 & 1,422,650,789 & 1,636,497,394 \\
         \\
         \hspace{-2mm} \textit{Published 2015-2022} & & & \\ 
         Number of Records & 22,609,069 & 27,620,472 & 76,836,191 \\ 
         Number of References & 786,437,547 & 1,035,750,923 & 840,730,834 \\ 
         \\
         \textit{Shared Corpus (2015-2022)} & & & \\
         Number of Records & 16,788,282 & 16,788,282 & 16,788,282 \\
         Number of References & 725,008,043 & 727,056,725 & 585,616,069 \\
\\
    \end{tabular}
    \caption{Sizes of databases and of the Shared Corpus dataset, with the number of references contained in each dataset}
    \label{tab:data-overview-size}
\end{table}

The versions of the WoS, Scopus and OpenAlex databases used in this study are as follows. The WoS and Scopus data are snapshots from five indexes of the WoS Core Collection (Science Citation Index Expanded, Social Sciences Citation Index, Arts \& Humanities Citation Index, Conference Proceedings Citation Index -- Science and Conference Proceedings Citation Index -- Social Sciences) starting from publication year 1980 and the Scopus database, both captured in April 2023. The OpenAlex database is the version released in August 2023, due in both cases to the versioning policy of our data host at FIZ Karlsruhe, the Leibniz Institute for Information Infrastructure.\footnote{\url{www.fiz-karlsruhe.de}}

Due to this discrepancy in version dates, we have decided to restrict the items in the Shared Corpus to those published on or before the 31\textsuperscript{st} December 2022 in order to mitigate any bias between the databases, and further refined this corpus to exclude records published before the 1\textsuperscript{st} of January 2015, so the Shared Corpus covers items from publication years 2015 to 2022 inclusive.

As the Scopus database mainly contains items from 1996 onwards (although since 2015, pre-1996 cited references and backfiles of major publishers have been added,\footnote{\url{https://blog.scopus.com/posts/breaking-the-1996-barrier-scopus-adds-nearly-4-million-pre-1996-articles-and-more-than-83}}) and WoS and OpenAlex have had no such restriction, to avoid bias in the computation of source reference counts and internal coverage we further restrict references to those items published between 1996 and 2022.

We include a section on articles published 2015-2022 in Table \ref{tab:data-overview-size} for all three databases, to illustrate the influence of the time restriction to the size of the Shared Corpus, and to give context to the DOI matching and deduplication work described in Section \ref{sec-data-dedup}.

In the Scopus and WoS databases, pre-computed total 'reference counts', are delivered by the data providers Elsevier and Clarivate, whereas 'source reference counts' are calculated for each record by our data provider FIZ Karlsruhe. Both databases are expected to contain all references of a given publication, regardless of whether they refer to items contained within or not contained within their databases, i.e. whether they are source and non-source references respectively, and without a fixed time restriction. 

In contrast, presently in OpenAlex there only exist source references (see the OpenAlex documentation \citep{priem2022}\footnote{\url{https://docs.openalex.org/api-entities/works/work-object\#referenced_works}} -- apart from a smaller segment of references to supposedly deleted items) and a `source reference count' has been calculated by FIZ Karlsruhe in our database. This fact explains the empty values for the average total reference counts in Table \ref{tab:computed_source_coverage}. We therefore have to relate this number to the source reference counts taken from WoS and Scopus.

\begin{figure}
\centering
\makebox[\textwidth]
    {\includegraphics[width=10cm]{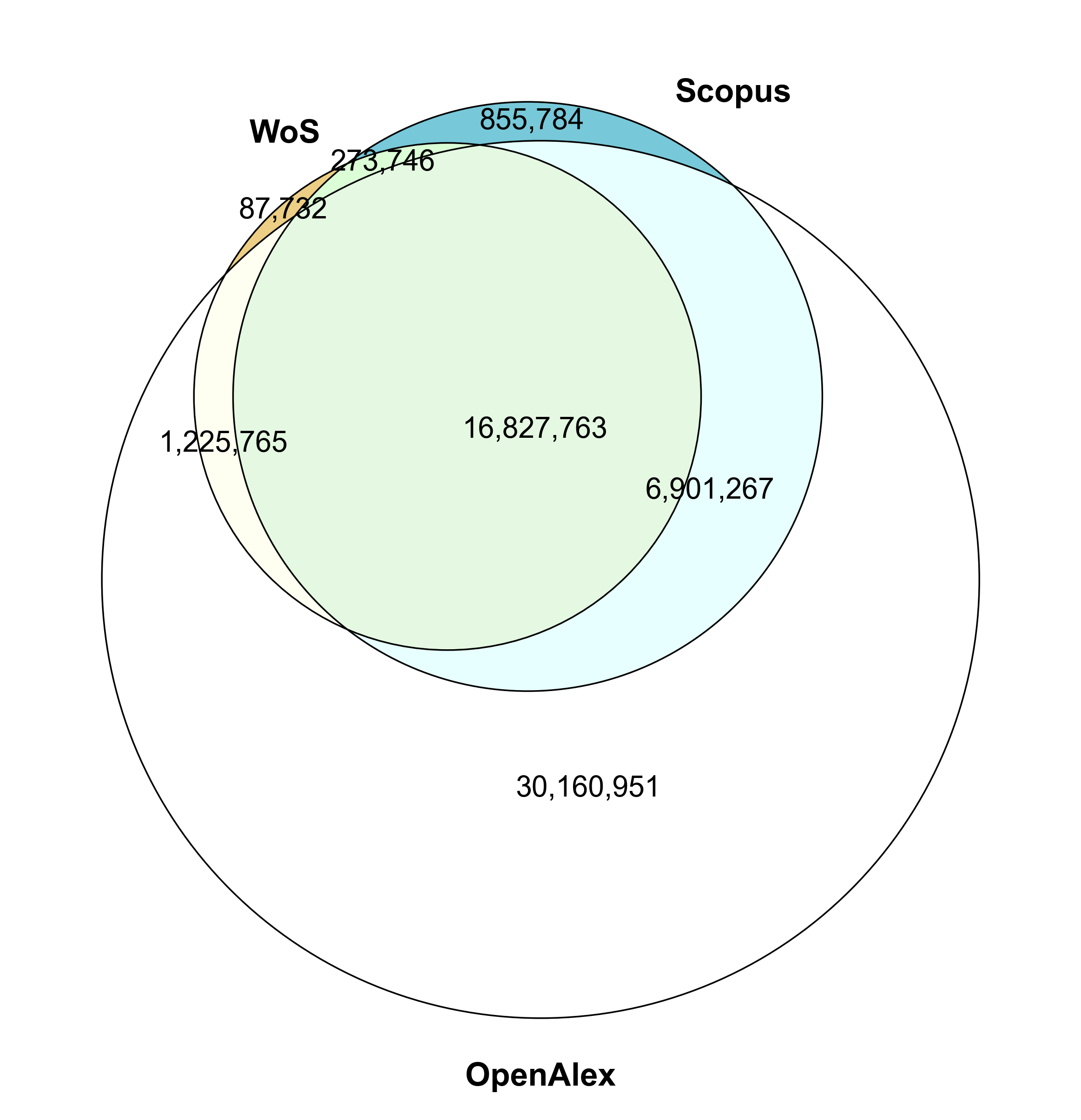}}
\caption{\label{fig:venn-2015}Venn diagram of the intersection sizes of unique DOIs based in each database on exact DOI match (without deduplication, i.e. cases of DOIs that have been assigned to multiple papers are now kept in the sets), for records published between 2015 and 2022.}
\end{figure}

In Table \ref{tab:data-overview-size} we provide a summary of the records available in each data source and in Figure \ref{fig:venn-2015} we provide a diagram of the intersections between the three data sources, based on exact matching of unique DOIs, over the entire corpus, and restricted to records published between 2015 and 2022. Additionally in Table \ref{tab:data-overview-size}, we provide information for the size of each corpus when restricted to records classified as `article' to demonstrate that this does not substantially decrease the relative scale of OpenAlex to WoS and Scopus. 

 To address the potential bias arising from the number of documents in the databases, we calculated the five-number summaries (median, standard deviation, maximum, minimum and inter-quartile range)
in addition to the mean values for the number references per article. Despite the considerable variation in the number of references per article and number of documents in the databases, we did not observe  substantial disparities in the distribution across the different data sources examined. Consequently, we have chosen to present the mean value.

It can be calculated from Table \ref{tab:data-overview-size}, that while the Shared Corpus, after DOI deduplication, contains 23.6\% and 25.6\% of all records in WoS and Scopus, and 6.9\% of those in OpenAlex, it contains 41.1\%, 35.8\% and 31.7\% of the references in the whole corpora of WoS, Scopus and OpenAlex respectively.

The Shared Corpus, after DOI deduplication, contains 74.3\% of the records in WoS published between 2015 and 2022, and 60.8\% of the records in Scopus published between 2015 and 2022 and 21.8\% of OpenAlex published between 2015 and 2022.

 To evaluate the reference and source reference coverage of WoS and Scopus against OpenAlex, we first used the reported reference counts and pre-calculated source reference counts as described in Section \ref{sec-method-data}.  
The average total reference count and source reference count was computed for: each database, for records marked as `article' (or comprising the document type `article' alongside other type markings, in the case of Scopus and WoS) and for the Shared Corpus resulting from the DOI match (publication years 2015-2022).

 These numbers were then checked by an independent calculation where the total number of references and records in each database were counted and the ratio was computed (`references per record'), as reported in Table \ref{tab:computed_source_coverage_discrepancies}. Then for the final results, queries were created to calculate and average the number of references with reference publication year 1996 to 2022, and the number of references that are linked to source items and publication years 1996 to 2022.

\subsection{DOI Match and Deduplication} \label{sec-data-dedup}
When constructing the Shared Corpus as described in Section \ref{sec-method-data}, we relied on the DOI as a unique identifier that we could use to combine the databases. This approach has its limitations, as explored in \cite{vieira_biblioverlap_2024} which highlights the distribution of non-existent or duplicate DOIs in each corpus may vary by subject in the WoS and Scopus databases.

We excluded records without a DOI and records where more than one publication item is attributed to the same DOI -- as we are virtually not able to decide which item is the correct one for a given DOI in the latter case. These duplicate records account for the removal of 39,481 publications (counted as distinct DOI) in addition to those resulting from the restriction to 2015-2022. This accounts for the difference between the size of the Shared Corpus and the nominal intersection of the three databases between 2015 and 2022. 

\subsubsection{Error Margins of the DOI Match}\label{sec-doi-dedup}
\begin{table}[ht]
    \centering
    \begin{tabular}{l|rrr}
          & WoS & Scopus & OpenAlex \\
          \hline
          \textit{Published 2015-2022} \\
          \hspace{2mm} DOIs with multiple Records & 7,177 & 76,891 & 11,074 \\
          \hspace{2mm} Records with a shared DOI & 14,376 & 282,893 & 22,158 \\
          \hspace{2mm} Records without DOI & 4,186,863 & 2,555,909 & 21,709,360 \\
    \end{tabular}
    \caption{A comparison of erroneous cases in the DOI match between databases}
    \label{tab:duplicate-records-comparison}
\end{table}
Records with a duplicate DOI or without a DOI were excluded from the DOI matching step in the construction of the Shared Corpus. 
In Table \ref{tab:duplicate-records-comparison}, which focuses on all publications in the three databases which are published between 2015 and 2022, it can be seen that Scopus has a significantly larger number of DOIs with multiple records associated with it.
Altogether,  OpenAlex has the greatest number of records without DOI, then WoS and Scopus.

As records without a DOI are not matched in our analyses there is a significant underestimation of the total size of the databases as portrayed in Figure \ref{fig:venn-2015}, similarly records which have a shared DOIs are counted once. 

Another reason for the exclusion of items in the DOI match, which at the same time restricts to publication years 2015-2022, is the fact that publication years are not always exactly the same between databases, possible due to differences in the handling of early access and print publication dates. We define the time restriction as applying to all three databases at the same time.

\subsection{Metadata Coverage}

To determine metadata coverage (as detailed in Section \ref{sec-results-open-metadata}), we also used the Shared Corpus as described at the beginning of Section \ref{sec-method-data}. Here, we restrict to publication items published in journals. For this purpose, the publication type categorisations of Web of Science and Scopus were used and the OpenAlex publications were assigned to these via the DOI comparison of the Shared Corpus, so that OpenAlex could be compared bilaterally with the other two databases.  We then specifically compared the coverage of abstracts, funding information, ORCIDs and Open Access (OA) status information by assessing whether items have (at least one) of these and aggregated by journal, that is, for each journal, a publication record was counted if the desired metadata property was available. In the case of OA, we counted the item if the OA status was not marked as \textit{closed}. We have normalised the journal title to lowercase to aggregate the items.
\section{Results} \label{sec-results-disc}

\subsection{Total and Source Reference Coverage}
Table \ref{tab:computed_source_coverage}, in a na\"ive averaging of the source reference count, leaves OpenAlex looking comparatively poor at 7.6 references per record to the 16.9 or 18.7 of WoS and Scopus (and well behind the other databases' average total reference count). However, when restricting to the 2015-2022 corpus shared by the three databases, OpenAlex proves competitive with a higher average source reference count than both WoS and Scopus. The fact that results vary greatly depending on the underlying corpus definition suggests that OpenAlex comprises of many publications with comparatively short reference lists which are not contained by the WoS or Scopus. When focusing on the comparison of the average total reference counts between WoS and Scopus, it initially appears that Scopus outperforms WoS, however when considering records marked as articles they perform more comparably. This trend continues when observing the Shared Corpus and the Shared Corpus with references from 1996 to 2022. Notably here the difference between the source reference count and total reference count decreases as the restrictions are added. 
The results suggest that Scopus still has a small disadvantage due to its initial indexing start in 1996.
Consequently, the slight advantage for OpenAlex is reversed when references are restricted to reference publication years 1996-2022, with Scopus outperforming OpenAlex, and WoS performing worst -- however, differences are very small.

\begin{table}[htbp]
    \centering
    \begin{tabular}{l|ccc}
        & WoS & Scopus & OpenAlex \\
        \hline 
        \hspace{-2mm}\textit{Whole Corpus} & & & \\
        Reported Average Reference Count & 24.765 & 31.254 & -- \\
        Pre-calculated Average Source Reference Count & 16.867 & 18.692 & 7.572 \\
        Internal Coverage & 68.1\% & 59.8\% & --\\
        \\
        \hspace{-2mm}\textit{Whole Corpus - Articles Only} & & & \\
        Reported Average Reference Count & 32.826 & 32.805 & -- \\
        Pre-calculated Average Source Reference Count & 22.442 & 20.230 & 8.134 \\
        Internal Coverage & 68.4\% & 61.7\% & --\\
        \\
        \hspace{-2mm}\textit{Shared Corpus (2015-2022)} & & & \\
         \textit{All References} &&&\\
        \hspace{2mm} Reported Average Reference Count & 43.185 & 43.320 & -- \\
        \hspace{2mm} Pre-calculated Average Source Reference Count & 33.416 & 33.363  & 34.863  \\
        \hspace{2mm} Internal Coverage & 77.4\% & 77.0\% & --\\
        \\
        \textit{References 1996-2022}&&&\\
        \hspace{2mm}Calculated Average Reference Count & 38.226  & 38.062 & -- \\
        \hspace{2mm}Calculated Average Source Reference Count & 31.207 & 33.359& 31.823\\
        \hspace{2mm}Internal Coverage & 81.6\% & 87.6\% & --\\
    \end{tabular}
    \caption{Comparison of the reference coverage available in each database, including the reported reference counts from the database providers, the pre-calculated source reference counts from FIZ-Karlsruhe, and our computed counts}
    \label{tab:computed_source_coverage}
\end{table} 
The internal coverage of OpenAlex cannot be computed for Table \ref{tab:computed_source_coverage} as it does not contain all references, respectively a total reference count. However, we can infer OpenAlex' internal coverage in the Shared Corpus by assuming either Scopus or WoS contain a definitive reference count. In this case, the internal coverage for the last segment (comprising the 1996-2022 restriction to reference publication years) for OpenAlex would be 83.2\% when related to WoS' total reference count, or 83.6\% when related to Scopus' reference count, notably these values lie between those of WoS and Scopus. We cannot perform the same analysis on all comparisons given the differing database sizes.

\subsection{Discrepancies between Reference Counts and Reference Data} \label{sec-discrepancies}
When comparing the reported and pre-calculated average total and source reference counts to an alternatively self-calculated ratio of all references to all publications, we came across discrepancies in Scopus and OpenAlex. In case of Scopus, reference counts reported by the provider do not always correspond to the actual references in the database, a phenomenon confirmed by Elsevier in informal communication as being caused by inconsistent supplier ingestions. In case of OpenAlex, some references refer to items that do not exist in OpenAlex, i.e. are deleted. The latter references are not included in the pre-calculated values. The discrepancies between both types of calculation can be seen in Table \ref{tab:computed_source_coverage_discrepancies}.

For further verification, we selected the publications in Scopus and OpenAlex where either the pre-calculated total `reference count' in Scopus and `source reference count' in OpenAlex were not equal to the respective number of entries in the databases' reference table. We then computed the averages of the reported/pre-calculated counts, and compared this to the ratio of references to publications while excluding in both cases the identified publications where reference count (in Scopus) or source reference count (in OpenAlex) do not correspond to the actual number of references. Once this has been done, the resulting averages then only differ at the 12th to 14th decimal place. We therefore conclude that for both databases discrepancies between reference counts and actual reference numbers are due to erroneous data. While in OpenAlex both our pre- and self-calculated source reference counts are consistent to our concept as we only count as source references those whose target items are actually in the database, the situation is more complicated in the case of Scopus: In our averages, we first use the reference counts supplied by the provider, which do not always match (but are probably more correct than) the references actually supplied, while in the last segment, where we calculate the count ourselves with references restricted to the 1996-2022 time window, we can only do this on the basis of the references actually supplied.

The detected discrepancies between both proprietary and open source bibliometric databases  should be considered when working with OpenAlex for bibliometric analyses - as averages of reference counts may differ significantly if the databases are not judiciously curated. We believe this discrepancy likely merits a deeper analysis in OpenAlex as new versions are released.

\begin{table}[!ht]
    \centering
    \begin{tabular}{l|ccc}
        & WoS & Scopus & OpenAlex \\
        \hline 
        \hspace{-2mm}\textit{Whole Corpus} & & & \\
        Ratio of References per Record & 24.765 & 30.979 & 7.592 \\
        Reported Average Total Reference Count & 24.765 & 31.254 & -- \\
        Reported Average Source Reference Count & 16.867 & 18.692 & 7.572 \\
        \\
        \hspace{-2mm}\textit{Whole Corpus - Articles Only} & & & \\
        Ratio of References per Record & 32.826 & 32.645 & 8.155 \\
        Reported Average Reference Count & 32.826 & 32.805 & -- \\
        Reported Average Source Reference Count & 22.442 & 20.230 & 8.134 \\
    \end{tabular}
    \caption{Discrepancies between Scopus and OpenAlex reported / pre-calculated reference counts and the ratio of references to records}
    \label{tab:computed_source_coverage_discrepancies}
\end{table} 

\subsection{Metadata by Journal} \label{sec-results-open-metadata}
\begin{figure}
\centering
\makebox[\textwidth]
    {\includegraphics[width=14cm]{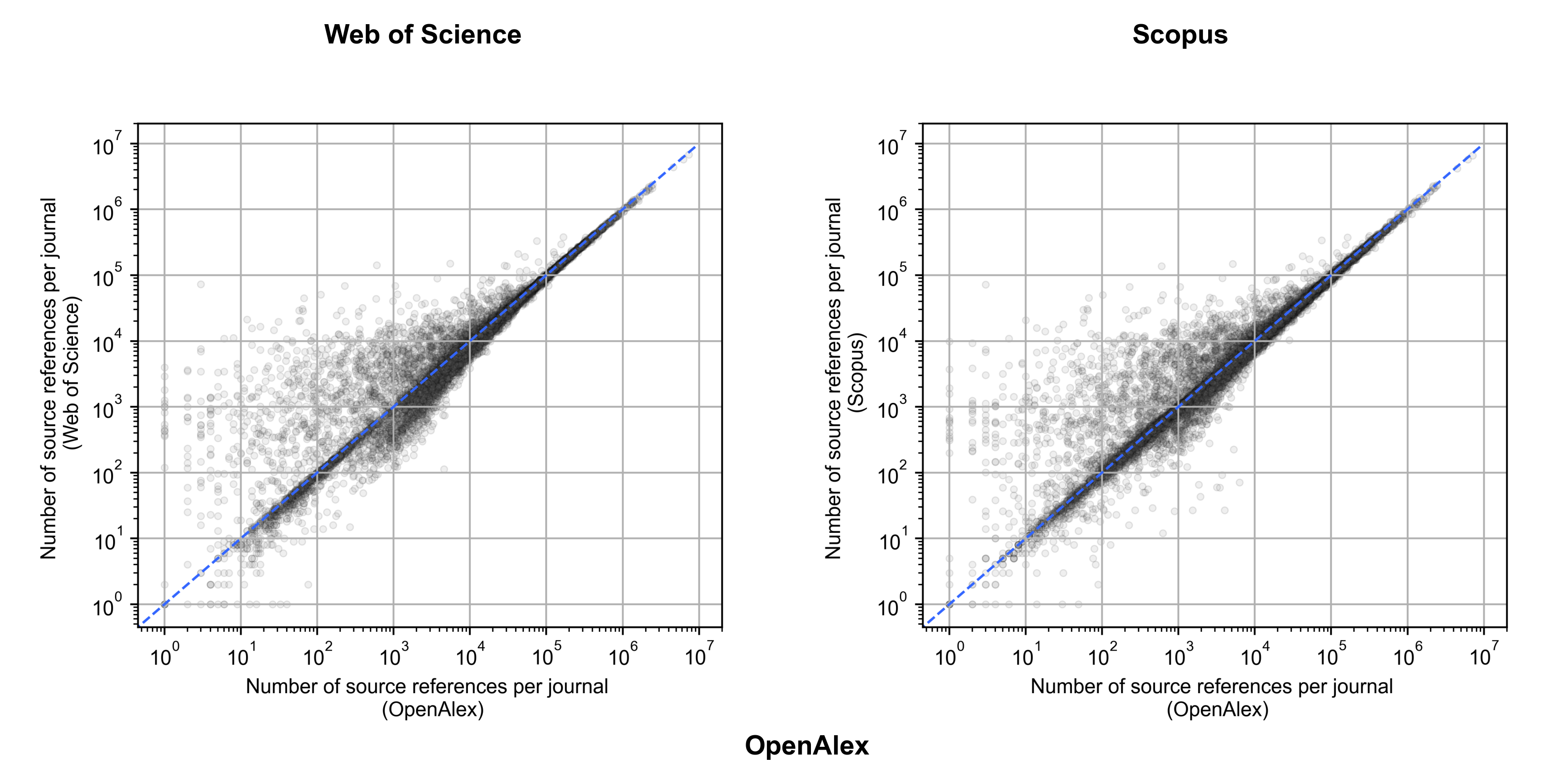}}
\caption{\label{fig:scatter_by_source_ref}Scatter diagrams of the count of source references per journal between OpenAlex and the Web of Science and Scopus.}
\end{figure}

\begin{figure}
\centering
\makebox[\textwidth]
    {\includegraphics[width=14cm]{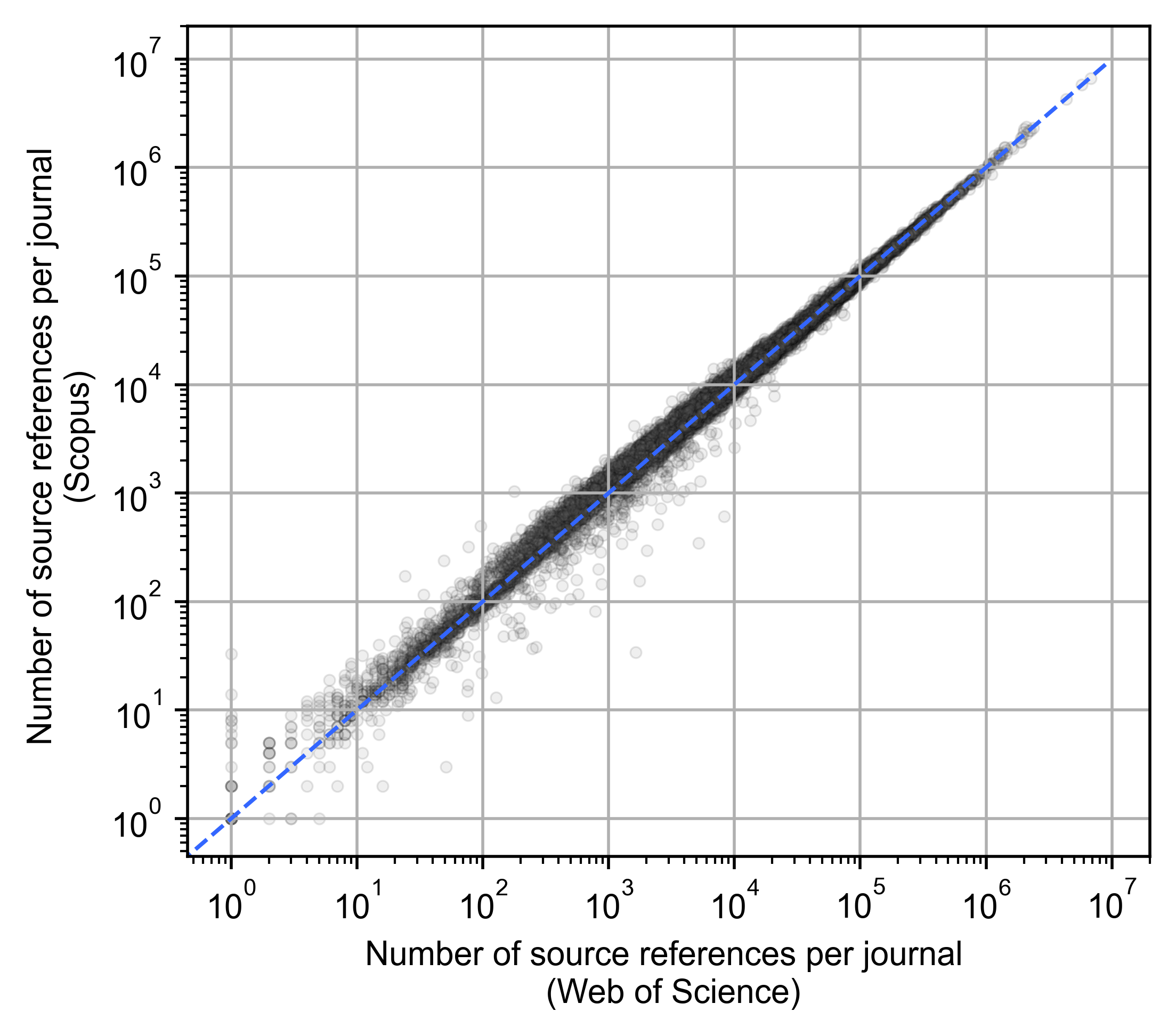}}
\caption{\label{fig:scatter_by_source_ref_wos_sco}Scatter diagrams of the count of source references per journal between OpenAlex and the Web of Science and Scopus.}
\end{figure}

Continuing the analysis of OpenAlex, WoS and Scopus, we then broke down the data by journal. Firstly in Figure \ref{fig:scatter_by_source_ref}, we compared the counts of source references in each journal in the WoS and Scopus to those in OpenAlex, spotting a fairly similar distribution in the two comparisons. Comparing these to Figure \ref{fig:scatter_by_source_ref_wos_sco}, we observe that the greater density under the $y=x$ line indicates OpenAlex is on average identifying slightly more source references in some journals, but the lesser density above the line indicates OpenAlex significantly undercounts on some journals as compared to WoS and Scopus.


\begin{figure}
\centering
\makebox[\textwidth]
    {\includegraphics[width=14cm]{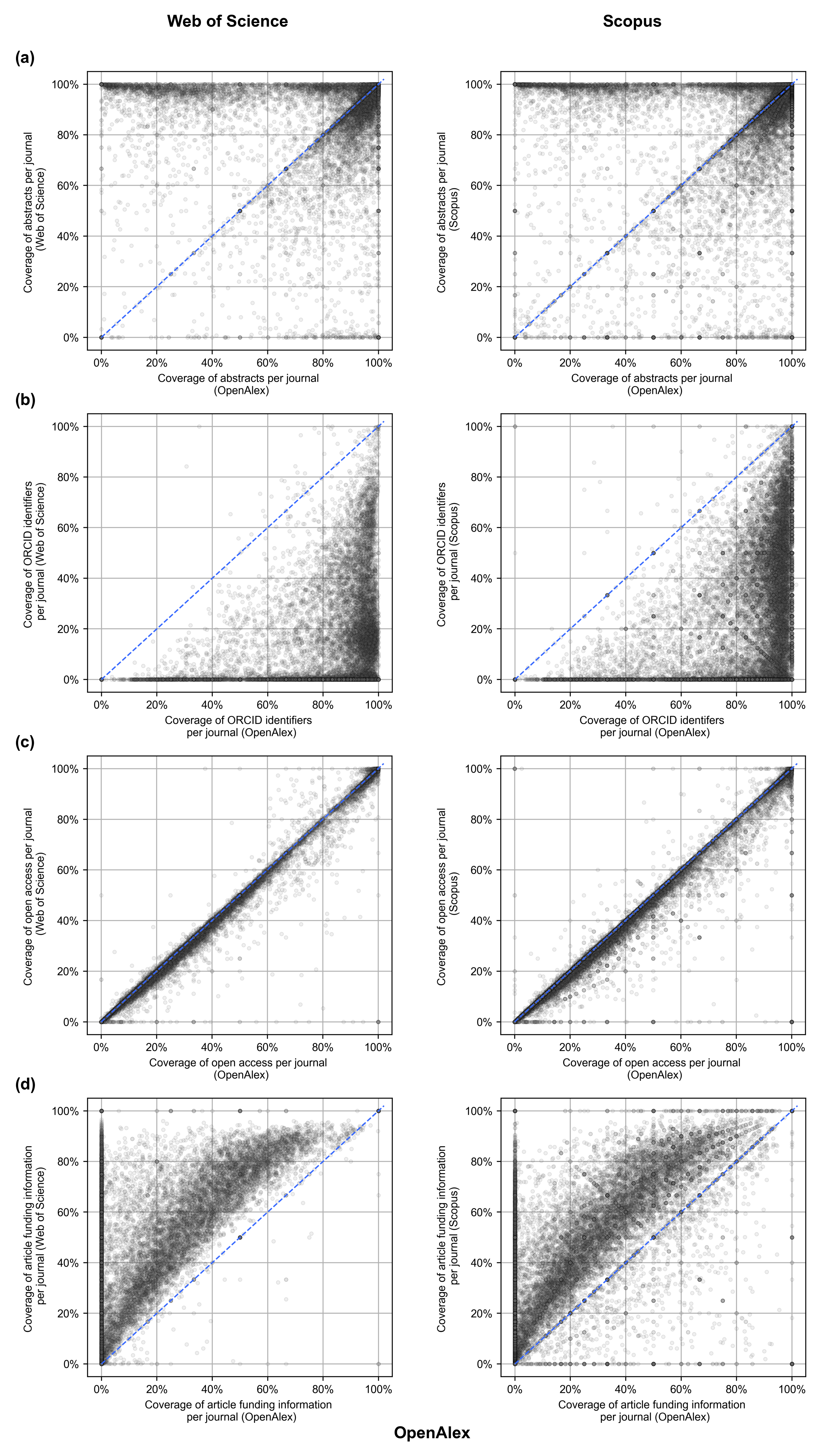}}
\caption{\label{fig:open_metadata}Scatter diagrams of the coverage of metadata per journal between OpenAlex and the Web of Science and Scopus.}
\end{figure}

 Figure \ref{fig:open_metadata} then highlights the metadata coverage analysis results between OpenAlex and the two proprietary data\-bases, WoS and Scopus, within the Shared Corpus. The x-axis represents OpenAlex, while the y-axis corresponds to WoS (left) and Scopus (right). The points represent the percentage coverage of the relevant indicator per journal.
 
The results indicate that OpenAlex depicts a different pattern compared to WoS and Scopus in terms of abstracts (Figure~\ref{fig:open_metadata}a), with the two proprietary databases having a higher overall availability of abstracts. In particular we note that there are concentrations near the top of the plot, indicating that the proprietary databases have full access to abstract information where OpenAlex has either partial or no access to this information.

Observing the top right of Figure \ref{fig:open_metadata}a, we see that the majority of journals reside in this area; in total, over 92\% of the articles in WoS and Scopus have abstract information, compared to a 87\% coverage of abstracts in OpenAlex, which implies a density in the top right hand corner which is not well indicated in the scatter diagrams.
Furthermore we see below the $y=x$ line that for some journals OpenAlex has a higher abstract coverage.
 
 In contrast, the ORCID coverage is more comprehensive in OpenAlex (Figure~\ref{fig:open_metadata}b). The proportion of articles in OpenAlex with at least one ORCID present is 92\%, and the proportion of articles with at least one ORCID in WoS is 16\% and in Scopus 32\%. However, upon inspection we discovered that OpenAlex performs a generous disambiguation of authors, resulting in a high ORCID coverage. In particular, some authors with Chinese names were observed to be linked to more than 10,000 publications.
 
 The distribution of open access information is more similar between the databases (Figure~\ref{fig:open_metadata}c), with a tendency slightly in favour of OpenAlex, suggesting an indexing lag of Unpaywall's open access status information in the WoS and Scopus data. The proportion of open access information in all three databases is around 49\%.
 
 Viewing Figure~\ref{fig:open_metadata}d shows that the availability of funding information on articles is better represented in WoS and Scopus than in OpenAlex. Notably, funding information associated with articles in over 4,100 journals can only be found in WoS and Scopus, which could indicate a lack of provision of funding information by some scholarly publishers for open databases such as OpenAlex and Crossref.

\section{Discussion}

This report demonstrates the source reference coverage in OpenAlex to be comparable to that in WoS and Scopus for comparatively newer records which lie in the intersection of all three databases, both in general and when restricting to references from 1996 onwards. On the one hand, this can be seen as an indicator of good quality bibliometric core data. On the other hand, OpenAlex does not have the highest internal coverage, although it is by far the largest database, so it would actually be plausible that higher proportions of the referenced publications are themselves part of the database. In this respect, the Scopus coverage policy seems to be a bit more effective. However, one possible factor could also be that a comparatively poorer reference-matching algorithm misses a noticeable amount of actual source references. 

 The vastly greater corpus of document records in OpenAlex, compared to WoS and Scopus, raises the question of what this additional content is, which is covered by OpenAlex but by neither established commercial provider. Our findings demonstrate what this content is not: it is not that part of the scientific literature which is referenced by items within WoS or Scopus. If that were the case, we would have found that OpenAlex internal reference coverage clearly exceeding that of the other to data sources in the Shared Corpus, because more references cited by those publications would be indexed by OpenAlex, but not WoS and Scopus. The substantially larger differences between the mean source reference counts of OpenAlex and the other two databases, if the entire databases and not the fixed comparison corpus are taken, also shows that the publications that are only in OpenAlex and not in the other two databases pull down the mean values due to their lower reference counts. They must therefore represent a different publication spectrum or have a significantly lower data quality. In any case, this suggests that OpenAlex should be limited to a core corpus if comparability of bibliometric analyses based on OpenAlex to WoS and Scopus is desired.
  
From Table \ref{tab:computed_source_coverage} it can be inferred that within the Shared Corpus, there are on average 6.4 to 6.2 references captured in the total reference count by WoS and Scopus (respectively) that OpenAlex does not capture in its source reference count. The fact that OpenAlex does not yet systematically include non-source references, as well as complete reference strings, limits the flexibility of using and exploring the data source: It does not allow researchers or bibliometric centers to apply their own reference matching algorithms or to analyse non-source references as such.

The study also revealed data errors in Scopus and OpenAlex. The reported figures for reference counts in Scopus do not correspond to the actual numbers of references in the database, and OpenAlex is inconsistent in its handling of references as it does not systematically comprise all non-source references, but references to some deleted source items. Similarly, we note that all databases, to a different degree, comprise cases where DOIs refer to multiple records--cf. \cite{franceschini2015errors}. We believe it merits further study and caution when replicating these computations. Another study by \cite{hauschke_non-retracted_2024}, in preprint at time of writing, indicates that data errors have been discovered in the "is\_retracted" field of OpenAlex for publications between 22 December 2023 and 19 March 2024, further highlighting the volatility of the metadata quality in OpenAlex.

 In summary, from an internal reference coverage perspective, OpenAlex as a source for citation data for studies of contemporary scientific output, is on par with commercial databases when restricted to a core corpus of publications similar to that of WoS and Scopus. However, its utility is hampered by not yet providing full cited reference data.

 Although metadata coverage relating to abstract information is lower than in WoS and Scopus, the share of records with abstracts in OpenAlex is nevertheless higher than in Crossref as stated in \cite{kramer_2024}. 
Kramer also notes that, at the time of writing, the large publishers Elsevier, Taylor \& Francis and IEEE did not openly share abstracts via Crossref. But OpenAlex also acknowledged legal issues, which resulted in the representation of abstracts as inverted index as well as in the removal of some abstracts.\footnote{\url{https://groups.google.com/g/openalex-users/c/ptFDD7qWvYw/m/kXWDG3o5BAAJ}}

Our analysis reveals that OpenAlex demonstrates a particularly high level of coverage for ORCID in comparison to WoS and Scopus. Over 90\% of articles in OpenAlex had been assigned at least one ORCID. However, we have observed that this percentage is somewhat excessive. Upon inspection, we discovered that in some cases ORCIDs were assigned to more than 10,000 records in our corpus, suggesting issues with OpenAlex's author disambiguation method. The authorships and author records have also been subject to updates by OpenAlex since data collection, including cleaning of author strings, syncing 17.9 million work records with Crossref, removing 3.9 million empty author records (authors with no works assigned to them) and updating author information fields.

In conclusion, our analysis of the metadata by journal highlights data collection and curation challenges for OpenAlex, having to collate information from both bibliometric and non-bibliometric sources\footnote{\url{https://help.openalex.org/hc/en-us/articles/24397285563671-About-the-data}} requires OpenAlex to perform disambiguation and standardisation between data sources, both challenging tasks, as well as deal with legal constraints in collecting and publishing academic works -- for example, the copyright of the abstracts. These challenges likely differ from those of WoS and Scopus in their collection and curation, but the similarity of the figures plotting OpenAlex against WoS and OpenAlex against Scopus demonstrate a stark difference between OpenAlex, and WoS and Scopus. Therefore we currently recommend caution when utilising OpenAlex for scientometric studies due to the volatility and data quality issues discussed earlier in this section.

\section{Limitations and Outlook} \label{sec-limitations-outlook}

We restate that our data is representative of late 2023, with the hitherto noted volatility of OpenAlex in the time since, this report may not be representative of the state of OpenAlex, and also the Web of Science and Scopus, at time of publishing.

A fundamental limitation of our study setting is the lack of ground truth--we do not analyse whether the reference counts provided by WoS and Scopus correspond exactly to the respective reference lists in the publications. 
However, we have checked in all three cases whether delivered and pre-calculated reference counts and delivered references correspond.

We also do not check the accuracy with which the databases match references to publications, which can be seen as the prerequisite for the internal coverage indicator we use. Some studies analyse the accuracy of the database matching algorithms either on the basis of manual sample evaluations and/or in comparison with their own algorithms for example, in \cite{Olensky2016}.

In a more extensive setting, an in-depth comparison of source and non-source references of each publication in a sample between the databases could provide indications of the extent to which the detected smaller differences can be explained by different coverage profiles or strengths and weaknesses of the matching algorithms.
A possible extension of our main methodological setting could analyse the internal coverage with respect to the disciplinary level and address the question to what extent OpenAlex has a better (or worse) coverage of non-English, regionally-oriented journals which might be relevant to some arts \& humanities and social sciences subjects, for example, and do not easily fulfil WoS curation criteria.

When studying ORCID availability, it must be noted that we did not check for the availability for all co-authors, but just if there was at least one ORCID present per article. It is important to conduct further analysis to confirm whether the author names and ORCIDs are accurately matched, given the observed phenomenon of a single ORCID being erroneously attributed to tens of thousands of articles. If this is not the case then this may demonstrate the ongoing challenge of author name disambiguation in bibliographic databases.

As discussed in Section \ref{sec-data-dedup}, some DOIs were found to have duplicate records assigned to them in each of the three databases, requiring us to deselect the 39,481 records from 2015-2022 which lay in the intersection of the three databases and had more than one record associated with the DOI in one of the databases from our Shared Corpus. 
A more detailed examination of duplicate DOIs may be merited, in particular with respect to Scopus (as demonstrated by Table \ref{tab:duplicate-records-comparison} 
and agreeing with findings in reported in \cite{vieira_biblioverlap_2024} on Scopus.) Similarly investigations into the distribution of duplicated, missing or incorrect DOI by record type between each database may be recommended for future research. 

Therefore as highlighted in \cite{vieira_biblioverlap_2024} and \cite{nikolic_open-source_2024}, where procedures for merging and deduplicating bibliographic datasets are explored, we note that DOIs are not perfect identifiers for combining datasets and often, due to duplicated, missing or incorrect DOIs, lead to the requirement for more demanding title and author name comparisons for generation of better quality datasets for future study.

Since the data collection effort of this study, at least 151 million references have been added to OpenAlex, as of the May 30\textsuperscript{th} 2024 snapshot of OpenAlex reportedly expanding the number of references by 7.61\% compared to the April 25\textsuperscript{th} 2024 snapshot. This and other ongoing efforts to improve data quality and expand data availability in OpenAlex indicate the need for future similar studies, and highlight the volatility of the database.

\section*{Declarations}

\bmhead{Funding}
This work was funded by the Federal Ministry of Education and Research via funding numbers: 16WIK2301B / 16WIK2301E, The OpenBib project \citep{schmidt_2024_13932928}.
We acknowledge support by Federal Ministry of Education and Research, Germany under grant number 01PQ17001, the Competence Network for Bibliometrics.

Jack Culbert and Philipp Mayr received additional funding by the European Union under the Horizon Europe grant OMINO – Overcoming Multilevel INformation Overload\footnote{\href{http://ominoproject.eu}{http://ominoproject.eu}} under grant number 101086321 \citep{holyst024}.

\bmhead{Prior Public Release} 

The findings presented in this paper were initially reported in a preprint on ArXiv \citep{culbert_reference_2024}. That version includes a longer and more extensive version of the results.

\bibliography{references}

\end{document}